# Recent developments in comprehensive analytical instruments for the culture heritage objects – A review

Yuanjun Xu, Zhu An[*], Ning Huang, Peng Wang, Ze He, Zihan Chen

*( Key Laboratory of Radiation Physics and Technology of Ministry of Education, Institute of Nuclear Science and Technology, Sichuan University, Chengdu, 610064, China )*

**Abstract:**

This paper introduces the necessity and significance of the investigation of cultural heritage objects. The multi-technique method is useful for the study of cultural heritage objects, but a comprehensive analytical instrument is a better choice since it can guarantee that different types of information are always obtained from the same analytical point on the surface of cultural heritage objects, which may be crucial for some situations. Thus, the X-ray fluorescence (XRF)/X-ray diffraction (XRD) and X-ray fluorescence (XRF)/Raman spectroscopy (RS) comprehensive analytical instruments are more and more widely used to study cultural heritage objects. The two types of comprehensive analytical instruments are discussed in detail and the XRF/XRD instruments are further classified into different types on the basis of structure, type and number of detectors. A new comprehensive analytical instrument prototype that can perform XRF, XRD and RS measurements simultaneously has been successfully developed by our team and the preliminary application has shown the analysis performance and application potential. This overview contributes to better understand the research progress and development tendency of comprehensive analytical instruments for the study of cultural heritage objects. The new comprehensive instruments will make researchers obtain more valuable information on cultural heritage objects and further promote the study on cultural heritage objects.

---

[*] e-mail: anzhu@scu.edu.cn (corresponding author)

**Contents**

**1. Introduction**

Cultural heritage objects can be considered as a symbol of human civilization and not only the complicated information on the ancient culture were expressed, but also the production activities of the ancients were reflected, witnessing the vicissitudes of history. It is necessary for us to investigate various cultural heritage objects for understanding the raw materials and manufacture technologies used, the custom and culture expressed, as well as providing the helpful information for the conservation and restoration of the cultural heritage objects. For example, the pigments and other materials used in ancient paintings, manuscripts and murals and so on were investigated for understanding the pigments and art techniques, the provenance and alteration pathway of the pigments, creation time, authenticity, the preservation state, and restoration methods [1-9]. Other archaeological artifacts such as the potteries, porcelains and so on are investigated for obtaining the information on the raw materials, firing temperature, production technologies and preservation state of the archaeological artifacts [10-25]. The studies of some unusual cultural heritage objects, e.g., carpentry works, bricks, India's Ellora caves and so on allow researchers obtain the information on the materials, production technologies, conservation of artwork and so on [26-39].

Given the highly value and historical significance of these cultural heritage objects, it is obvious that any destructive investigation should be avoided. The nondestructive analytical techniques, e.g., XRF, RS, Fourier transform infrared spectroscopy (FTIR) and so on are usually used to study cultural heritage objects [40-46]. Only one type of information on the sample can be provided by one common analytical instrument. If ones hope to obtain more information on these cultural heritage objects, several different instruments would be used [47-59]. The combined use of several different instruments is called multi-technique method. It is widely used in the research field of cultural heritage objects since different types of information contribute to better understand cultural heritage objects and make the identification more complete and unambiguous [60-73]. Using the multi-technique method for the investigation of cultural heritage objects has many advantages, which are confirmed by a large number of applications of the multi-technique method in the research field of cultural heritage objects [74-82]. But for some special cultural heritage objects that are prohibited for sampling and moving, the multi-technique method is no longer the best choice. A comprehensive instrument where different types of analytical techniques are integrated into a single device is more suitable for the investigation of these special cultural heritage objects. Such comprehensive analytical instruments can guarantee that different types of information are always obtained from the same analytical point on the surface of cultural heritage objects, which may be crucial for some situations where the surface is inhomogeneous or ones are only interested in specific area on the surface of cultural heritage objects. Moreover, a mobile comprehensive instrument makes it possible to study the cultural heritage objects that prohibit for sampling and moving in situ. At present, the XRF/XRD and XRF/RS comprehensive instruments are the most commonly used analytical tools in the research field of cultural heritage objects. The XRF/XRD comprehensive analytical instrument means that the XRF and XRD analytical techniques are integrated into a single device and the elemental and crystalline information can be obtained simultaneously for the same sample position. The element information obtained by XRF can offer the complementary information for the identification of crystalline phase. When the crystalline phase cannot be unambiguously determined by the XRD information, the elemental information can help determine the specific crystalline phase. Conversely, the specific elemental information can also be verified by the crystalline phase information. The XRF/RS comprehensive analytical instrument means that the XRF and RS analytical techniques are integrated into a single device. The XRF and RS measurements can be implemented simultaneously for the same sample position. The elemental and molecular structural information are the complementary information for each other and they are helpful for the unambiguously identification of the sample. With the advantages of comprehensive analytical instruments, they are used more and more frequently in the research field of cultural heritage objects. Meanwhile, these comprehensive analytical instruments are developed and improved continuously with the development of component, detector, electronic and so on. Two types of information can be obtained by these comprehensive analytical instruments, but researchers always want to obtain more information on the samples, especially in studying some cultural heritage objects that are composed of complicated mixtures. The comprehensive analytical instrument that integrated more analytical techniques into a single device can provide more valuable information and make it possible to identify these cultural heritage objects completely and unambiguously. However, until now, there is not any comprehensive analytical instrument available that three or more analytical techniques are integrated into a single device due to the limitation of structure, component, optical path, weight, size and so on. Hence, it is necessary to develop such a powerful comprehensive analytical instrument for the investigation of cultural heritage objects.

This paper will focus on an overview of comprehensive analytical instruments for the cultural heritage objects. We will review and discuss these comprehensive analytical instruments used in the investigation of cultural heritage objects. This paper will also introduce the recent development of the comprehensive analytical instrument for which different types of analytical techniques are integrated into a single device. It will be shown that the powerful comprehensive instrument as the analytical tool will be the latest development tendency in the research field of cultural heritage objects.

2. **Comprehensive analytical instruments**

With the development of components, detectors and other electronic techniques, analytical instruments that integrated only one technique have already been commercialized. But a common analytical instrument cannot satisfy the need of the investigation due to the complexity of cultural heritage objects. Then the multi-technique method is used by researchers to investigate these cultural heritage objects for more complete characterization. Although there are many merits for the use of multi-technique method, some disadvantages cannot be avoided. For example, all analytical results obtained by different types of analytical instruments cannot be guaranteed from the same analytical point on the cultural heritage objects. This influences the accurateness of analytical results; secondly, the analyses for the cultural heritage objects performed by several instruments one by one will cost more time and reduce the analysis efficiency; thirdly, more analytical instruments are needed, which increases human and financial costs; fourthly, it is not convenient to carry several instruments to the field when needing to study the cultural heritage object in situ. However, these disadvantages of the multi-technique method can be avoided by using a mobile comprehensive instrument that integrates two or more analytical techniques into a single device. Therefore, the comprehensive instruments that integrated two or more analytical techniques into a single device are more suitable for the investigation of cultural heritage objects than the multi-technique method with several individual instruments. A comprehensive analytical instrument can be used to obtain complementary information or even the complete results and retain the merits of the multi-technique method. It can be known that the analytical techniques, such as XRF, XRD, RS and so on are widely used in the research field of cultural heritage objects. In recent years, two of them are usually integrated into a single device to become a comprehensive analytical instrument. These comprehensive analytical instruments can be classified into different types according to the different analytical techniques that are integrated into a single device. For example, the comprehensive analytical instruments where the XRF and XRD analytical techniques are integrated into a single device are named as the XRF/XRD comprehensive instrument [83]. They can be used to investigate the cultural heritage object for obtaining the elemental and crystalline information for the same sample positions at the same time. The XRF/XRD comprehensive instrument not only combines the advantages that use both techniques separately, but also avoids the disadvantages of using both techniques individually. Similarly, the instruments where the XRF and RS analytical techniques are integrated into a single device are named as the XRF/RS comprehensive instrument [84]. The elemental and molecular structural information of the sample can be obtained simultaneously for the same sample positions. Although there are some other types of comprehensive analytical instruments, such as RS/LIBS [85], RS/FTIR [86] comprehensive instruments where other two different analytical techniques are integrated into a single device, some of them (e.g., RS/LIBS) are not suitable for the study of the cultural heritage objects because of the particularity that cultural heritage objects are precious and irreproducible and any destructive analyses should be avoided. The XRF/XRD comprehensive instruments are often used for the investigation of cultural heritage objects. In 2014, the available state-of-the-art analytical instrument was used by Van de Voorde et al [87] to investigate the "Mad Meg" painting by Pieter Bruegel the Elder. The portable XRF/XRD comprehensive instrument made it possible to study in situ and obtained the comprehensive characterization results for the painting. Hitherto, the trend that more and more research groups attempt to use comprehensive analytical instruments for the investigation of culture heritage objects is becoming common. Recently, a new instrument that integrated XRF and XRD analytical technique was used by Gomez-Moron et al [88] to characterize the pigments of vaults of the kings in the Alhambra located in Granada, Spain to identify the chemical elements and crystalline phase compositions present in the areas under study. The results obtained by the XRF/XRD comprehensive instrument allowed the complete characterization of the pigments. The XRF/RS comprehensive instruments are also very useful for the investigation of cultural heritage objects. The RS and XRF are widely used as complementary techniques, which are well suitable for the investigation of cultural heritage objects since they are nondestructive, sensitive and efficient and the measurement can be carried out in situ. This type of comprehensive analytical instrument is therefore becoming popular in the research field of cultural heritage objects. For instance, a new mobile XRF/RS instrument was used by Andrikopoulos et al [84] for the investigation of the Byzantine icons. In addition, Guerra et al [89] developed a portable XRF/RS instrument that equipped with a digital optical microscope for in-situ analysis of culture heritage objects that are prohibited for sampling and moving. The complete analytical results demonstrated the capability of the portable XRF/RS instrument. In fact, these comprehensive analytical instruments are becoming the most important analytical tool in the research field of cultural heritage objects. Therefore, the categories, characteristics and applications of comprehensive analytical instruments will be discussed and summarized in the following sections. A new comprehensive analytical instrument prototype that integrates three types of analytical technique into a single device will be presented, and its preliminary application will be shown.

**2.1 XRF/XRD instruments**

The XRF/XRD instruments can be classified to different types according to the different X-ray source, structure design and detectors used. Both the X-ray source based on synchrotron or X-ray tube can provide X-rays to these XRF/XRD instruments, but the use of X-ray tube is more widely because it can be obtained easily and it can provide X-rays to portable instruments. Although the X-ray source based on synchrotron is only suitable for the desktop instruments, it also plays an increasing important role in the study of archaeometry and the conservation of cultural heritage objects since the X-ray source based on synchrotron can provide higher quality X-rays than X-ray tube. For instance, a paper that used the synchrotron-based μ-XRF/XRD instrument as well as other techniques including traditional XRD, XRF and so on to investigate the pictorial materials was published by Herrera et al [90]. In their paper, the advantages of synchrotron-based μ-XRF/XRD instrument were corroborated by comparing with those results obtained from other traditional instruments. As the X-ray source based on synchrotron is not suitable for the in-situ analysis, its use is limited. Moreover, it is not easily available
8

to the most of research groups, which further limits the applications of the X-ray source based on synchrotron. As a result, these comprehensive analytical instruments based on X-ray tube are mainly discussed in this paper. A portable XRD equipped with XRF spectrometer was constructed by Uda et al [91] for the investigation of cultural heritage objects. The performance of the improved portable instrument was confirmed by the investigation of the bronze mirror from the Eastern Han Dynasty, a stupa and its pedestal from the Heian Period. The authors mentioned that the existing XRD instrument equipped with XRF spectrometer should be improved in near future by designing a two dimensional scanning stage and developing the software for the quantitative-analysis. Pifferi et al [92] presented a portable comprehensive analytical instrument where the XRF and XRD measurements can be implemented simultaneously and it was described for the aspects of the hardware, software and the testing experiment. This instrument was equipped with the Theta-Theta horizontal protractor to optimize the stability and reliability of instrumental structure so that the accurate and reliable results can be obtained when the culture heritage objects are analyzed in situ. Due to the light weight of the instrument, it was particularly suitable for analyzing the cultural heritage objects that cannot be moved and sampled. Of course, most impressive of all was the software of the instrument because of the easy operation in the instrumental control and the data collection, as well as the complete and fast data analysis. In order to evaluate the performance of the instrument, the NIST Si sample, a laurel leaf and a Carrara marble sample were analyzed, respectively. Shen [93] also used a portable XRF/XRD instrument to investigate the gemstones. The author considered that the instrument can be applied to the cultural heritage and gemological fields after the hardware development and software updating. Nakai et al [94] wrote a compelling overview in which the characteristics of six types of XRF/XRD instruments were described in detail, and the merits and demerits of each instrument were compared to each other carefully for showing the improved aspects of the XRF/XRD instruments. Abundant practical applications of these XRF/XRD instruments were presented, which demonstrated the potential of these XRF/XRD instruments in the research field of cultural heritage objects. Besides the comprehensive use of the XRF and XRD, the joint use of XRF and RS was also advised by the authors. In the case of X-ray source based on X-ray tube, these XRF/XRD instruments can be classified to two categories according to the difference of instrument structures. One type of instrument is the XRF/XRD instruments based on transmission configuration, the other type of instrument is the XRF/XRD instruments based on reflection configuration. The components, characteristics and applications of different types of XRF/XRD instrument will be discussed in the next sections.

**2.1.1 XRF/XRD instruments based on transmission configuration**

The XRF/XRD instruments based on the transmission configuration mean that the X-ray tube and the detector locate at the different sides of the sample. The X-rays irradiate the sample from one side while the detector detects the fluorescence and diffraction information emitted by the sample from the opposite side. The structural schematic diagram is shown in Fig. 1. The diffraction and fluorescence data can be obtained from the X-rays detected by a charge-coupled device (CCD) detector. Several XRF/XRD instruments based on transmission configuration are now in commercial production [95], but this type of XRF/XRD instrument is not suitable for the nondestructive investigation of cultural heritage objects since the samples to be characterized are in the form of powder. An instrument named as CheMin is a typical representative of XRF/XRD instrument based on transmission configuration and it was developed for the NASA's planetary XRF/XRD instrument for the Mars science laboratory [96]. The CheMin is mainly composed of an X-ray tube, a CCD detector and the sample holder. This type of XRF/XRD instrument is widely used by geologists for the in-situ mineral analysis and the results obtained by the instrument have proved that the CheMin is very useful for conducting fieldwork because geologists are able to test their hypothesis when in the field and alter their collection or analysis strategy based on the field data [97]. Another similar XRF/XRD instrument based on transmission configuration, i.e., Terra, was developed by inXitu Inc. for the in-situ mineral investigation [94]. It evolved from the CheMin XRF/XRD instrument. A Terra XRF/XRD instrument was used by Stern et al [98] during the AMASE 2010 campaign to analyze the fine grained (<150 μm) powders. The elemental and mineral compositions of samples were successfully identified by the Terra. In addition, Cornaby et al [99] presented a paper that used a breadboard XRF/XRD instrument based on transmission configuration for testing some of the key components involved in an XRF/XRD instrument design, and then using this XRF/XRD instrument to analyze the rocks and minerals in 2001. The instrument was constructed at MOXTEK, Inc. and the instrument's construction has been fund by NASA. The authors compared the capabilities of the front-side-illuminated (FSI) and back-side-illuminated (BSI) CCDs, and the results showed that the FSI CCD is more efficient than the BSI CCD in capturing single events, while the BSI CCD has a larger energy detection range, particularly in low-energy region than FSI CCD. At present, all of XRF/XRD instruments based on transmission configuration are destructive analytical tools. Therefore, this type of XRF/XRD instrument is more suitable for the geological research field or other fields that allow the destructive analyses, while it is not suitable for the investigation of precious cultural heritage objects.

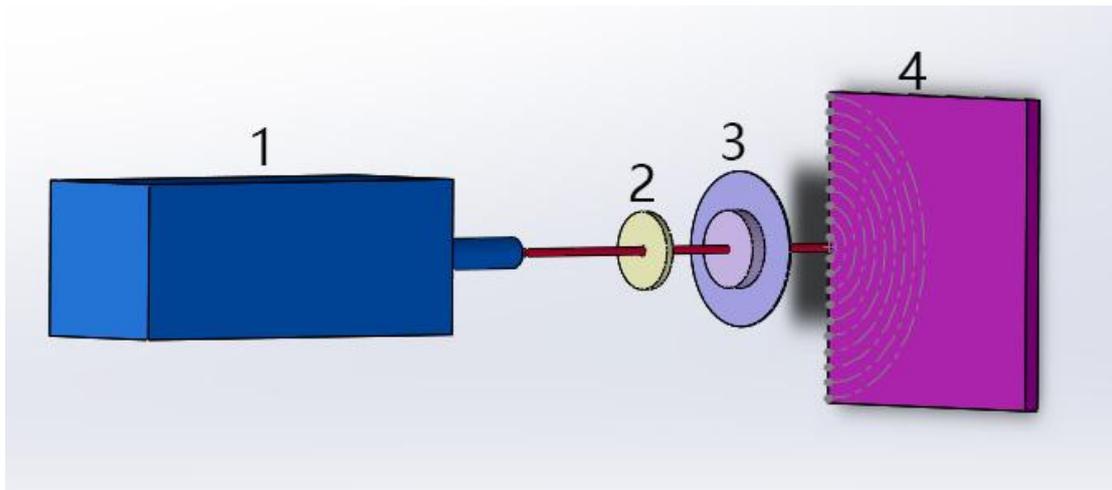

Fig. 1. The structural schematic diagram of XRF/XRD instrument based on transmission configuration

(1. X-ray tube, 2.Pinhole collimator, 3. Sample holder, 4. CCD detector).

**2.1.2 XRF/XRD instruments based on reflection configuration**

The XRF/XRD instruments based on reflection configuration refer to those where both the X-ray source and the detector system locate at the same side of the sample under study. The major difference of various XRF/XRD instruments based on reflection configuration is the detector used. Some XRF/XRD instruments employ a conventional protractor configuration in which the fluorescence and diffraction data are obtained by scanning the detector, e.g., SDD or Si-PIN, and/or X-ray source. Other XRF/XRD instruments make use of two-dimensional positive-sensitive detectors such as CCD or imaging plate (IP) [94]. Specifically, the XRF/XRD instruments based on reflection configuration can be further classified to five different types according to the type and the number of detectors and the analysis principles.

The first type of XRF/XRD instrument based on reflection configuration is mainly composed of an X-ray tube, a detector, i.e., Si-PIN or SDD detector and a Theta-Theta protractor. This type of XRF/XRD instrument is characterized by a combination of energy dispersive XRF with goniometry-based XRD data collection. And only a detector is used to collect the fluorescence and diffraction information simultaneously. The structural schematic diagram is shown in Fig. 2. The advantages of this type of XRF/XRD instrument are that the components of detector system are simpler, the instrument structure are more compact and have higher resolution and so on. But there are some disadvantages for this type of XRF/XRD instrument, such as troublesome mechanical movements, longer measurement time, complicated data processes and so on. Although the disadvantages will impact the development of this type of XRF/XRD instrument, the study and application of such XRF/XRD instrument are not impeded because of its advantages. For instance, De Voorde et al [100] described a new, commercially available, mobile XRF/XRD instrument that applied the Theta-Theta protractor. This attractive instrument is lighter and more compact comparing with other similar instruments due to the fact that it is composed of a miniature X-ray tube and a Si-PIN detector. This new XRF/XRD instrument could operate in two modes in which one is only in XRF mode while the other is in XRF and XRD mode where the XRF spectra could be acquired simultaneously when obtaining the XRD spectra. The components of the new instrument, the operation modes, the optimal selection of the pinholes, slits and the distance between the sample and the detector were described by the authors in detail. Several measurements were performed for calibrating the new instrument before identifying the chemical elements and crystalline phases of culture heritage objects in practice. As some culture heritage objects are prohibited for sampling and moving, the new XRF/XRD instrument is very suitable for being taken to the field and analyzing the culture heritage objects in situ because of its portability. In order to corroborate the potential of the new XRF/XRD instrument, the pigments and the lead alloyed printing letters were investigated. The results obtained by the new XRF/XRD instrument confirmed that the main crystalline phases present in various cultural heritage objects could be identified by the instrument non-destructively, especially for the cultural heritage objects that were not suitable for being analyzed by the traditional instruments. The paper [100] implied the importance of the new instrument where two different analytical techniques were integrated into a single device for obtaining more complete and reliable compositions information. Agresti et al [101] used a similar portable XRF/XRD instrument to investigate three bronze figurines from the Egyptian Museum of Florence. Impressively, the important chemical elements and crystalline phase information were obtained by the portable XRF/XRD instrument and the exhaustive characterization of the bronze figurines was realized by the portable XRF/XRD comprehensive analytical instrument.

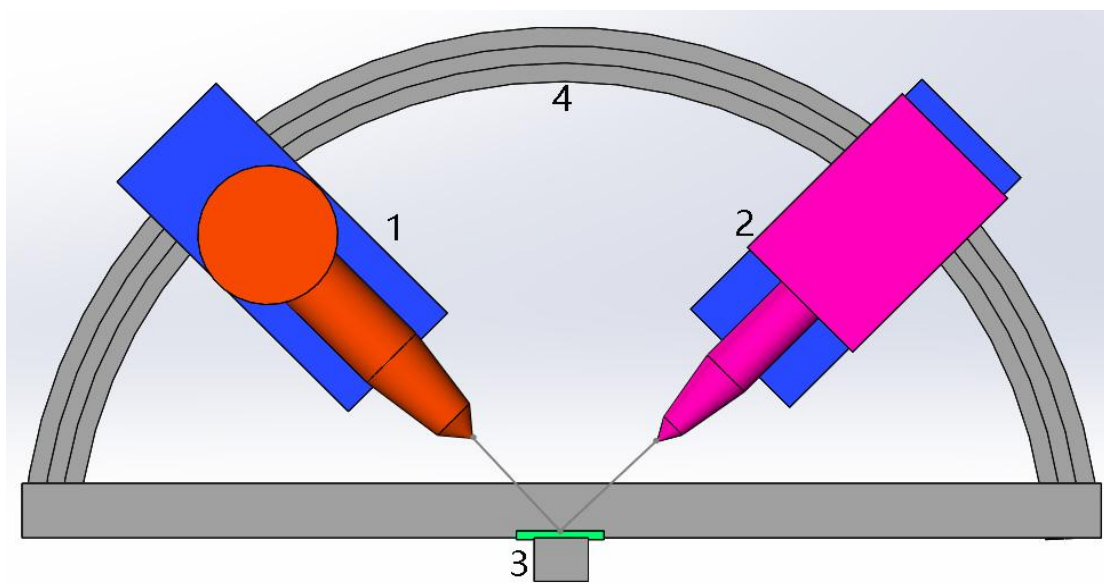

Fig. 2. The structural schematic diagram of first type of XRF/XRD instrument (1. X-ray tube, 2. Si-PIN or SDD detector, 3. Sample, 4. Theta-Theta protractor).

The second type of XRF/XRD instrument based on reflection configuration is equipped with an X-ray tube and a Si-PIN or SDD detector. The difference between the second and first type of XRF/XRD instruments consists in the different analysis principles, i.e., the energy dispersive X-ray diffraction (EDXRD) in the second type of XRF/XRD instrument replaces the angle dispersive X-ray diffraction (ADXRD) in the first type of XRF/XRD instrument. Since the energy dispersive X-ray diffraction utilizes the polychromatic light of different X-ray energies to identify the crystalline phases present in the sample, no protractor is required and the positions of X-ray tube and the detector are fixed. The XRF and XRD measurements can be made simultaneously, as well as the fluorescence and diffraction information can be collected by a Si-PIN or SDD detector. The structural schematic diagram is shown in Fig. 3. The advantages of this type of XRF/XRD instrument are the simpler detector system, more compact structure, simpler structural design, faster data acquisition, wider accessible region of the reciprocal space and no mechanical movements. It seems to be an ideal XRF/XRD instrument. However, the disadvantages of this type of XRF/XRD instrument are also prominent. The separation and interpretation of the data are very complicated due to the fact that the characteristic and diffraction peaks appear in the same spectrum, sometimes overlapping. In addition, the lower resolution and sensitivity are also the limitations. At present, the disadvantages of such XRF/XRD instruments cannot be avoided, but it deserves to further develop because of its prominent advantages. For example, Cuevas et al [102] reported how to improve the limitations of such XRF/XRD instrument. An XRF/XRD instrument that the merits of shorter acquisition time and higher energy penetration in EDXRD apparatus and the merit of higher inter-planar distance resolution in ADXRD apparatus were combined was built by Cuevas et al [102]. The diffraction and fluorescence peaks can be distinguished by changing the angle between the X-ray tube and detector manually. Although the authors gave a way to improve the limitations of such XRF/XRD instrument, the problems have not been solved completely. Hence, such XRF/XRD instrument is seldom used in the study of cultural heritage objects.

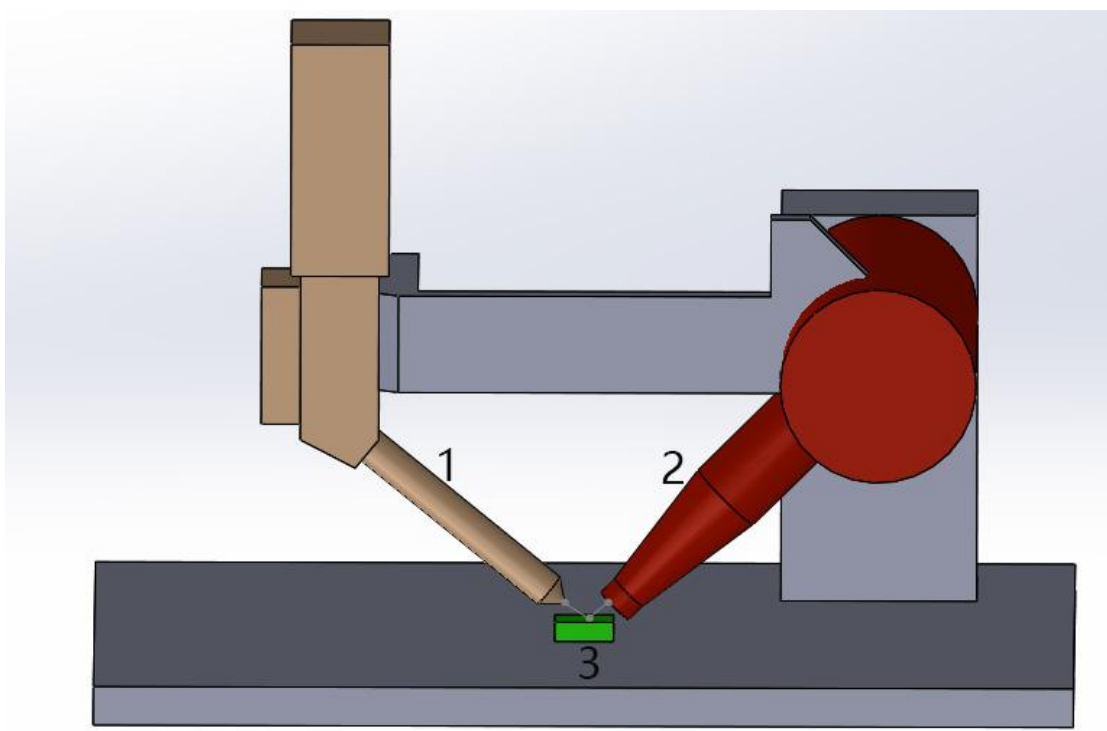

Fig. 3. The structural schematic diagram of second type of XRF/XRD instrument (1. X-ray tube, 2. Si-PIN or SDD detector, 3. Sample).

The third type of XRF/XRD instrument based on reflection configuration is mainly composed of an X-ray tube, a two-dimensional (2D) CMOS detector, a Si-PIN or SDD detector and a Theta-Theta protractor. The structural schematic diagram is shown in Fig. 4. This type of XRF/XRD instrument has the advantages of higher resolution, more stable, more accurate analytical results, easier data processing and so on. Meanwhile, they have the disadvantages of complicated structural design, troublesome mechanical movements and longer measurement time. Such XRF/XRD instruments are reliable and the analytical results are often very accurate because of the wider 2θ range, the higher angle resolution. The fluorescence and diffraction information are collected by different detectors, respectively, hence, the data processing becomes easier. Lutterotti et al [103] presented such an XRF/XRD instrument. An Amptek X123 detector and an Xpad 2D detector are used to obtain the chemical elements and crystalline phase information, respectively. The X-ray tube and the Xpad 2D detector operated in a scanning mode, while the Amptek X123 detector remained at a fixed position close to the sample. Considering the merits of such type of XRF/XRD instrument, it was used to investigate some archaeological finds from an alpine pastoral enclosure in Val Pore, Italy [103]. The analytical results obtained from the study corroborated the usefulness and validity of the instrument in the analyses of culture heritage objects [103].

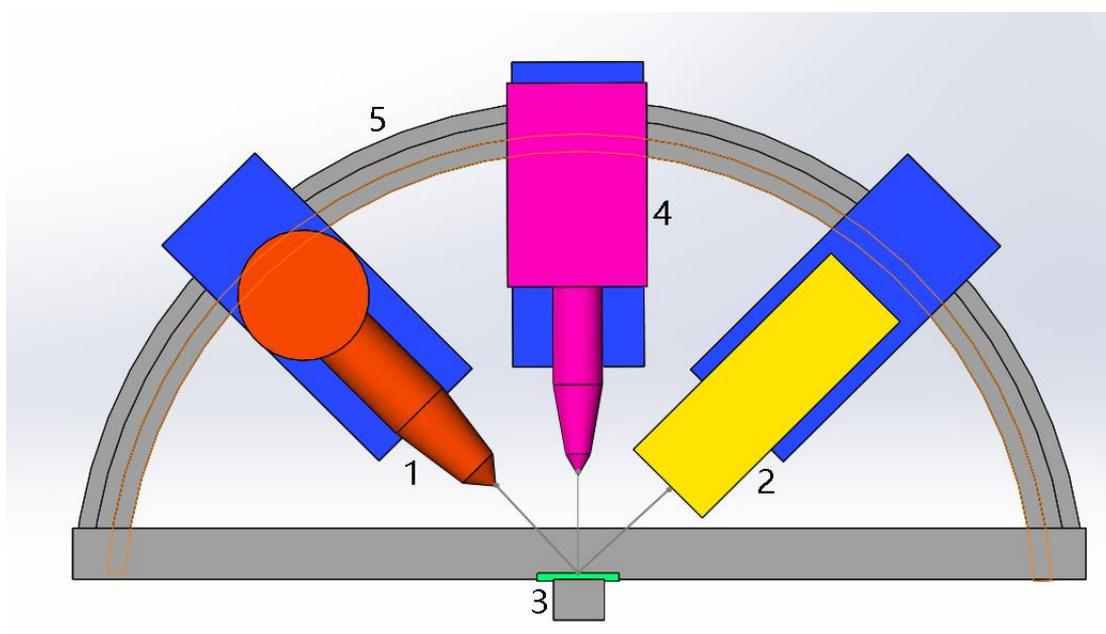

Fig. 4. The structural schematic diagram of third type of XRF/XRD instrument (1. X-ray tube, 2. Two-dimensional CMOS detector, 3. Sample, 4. Si-PIN or SDD detector, 5. Theta-Theta protractor).

The fourth type of XRF/XRD instrument based on reflection configuration that equipped with a two-dimensional positive-sensitive detector avoids the disadvantages of the first and third types of XRF/XRD instruments. There is no complicated structural design and no need to move the X-ray source and the detector. Such XRF/XRD instruments are composed of an X-ray tube, a Si-PIN or SDD detector and a two-dimensional detector, i.e., IP. The Si-PIN or SDD and two-dimensional detectors are used to collect the fluorescence and diffraction information, respectively. Although the two-dimensional detector collects two-dimensional diffraction image, it can be turned into a classical "intensity versus 2θ angle" curve by the free software FIT-2D. The structural schematic diagram is shown in Fig. 5. Such XRF/XRD instruments have many advantages, e.g., simpler structural design, faster data acquisition, easier data processing and interpretation and no mechanical movements. Although there are many advantages, such XRF/XRD instruments have a crucial disadvantage, i.e., needing an additional data reader for IP detector, which makes data collection become troublesome and increases the complexity of detector system. Considering many advantages of such XRF/XRD instruments, they are also widely applied in the analyses of cultural heritage objects. For instance, Duran et al [104] reported the characterization of illuminated manuscripts by such a portable XRF/XRD instrument in which a SDD and an IP detector were used to collect the fluorescence and diffraction information, respectively. Later on, Duran et al [105] reported another paper that used the portable XRF/XRD instrument for the investigation of the pigments of the Roma and Arabic murals. Another similar XRF/XRD instrument was used by Eveno et al [106] to investigate a variety of artworks for testing the quality and limits of the XRF/XRD instrument. The accuracy of the analytical results obtained by the XRF/XRD instrument was demonstrated by comparing with the results of traditional laboratory analytical instruments. The prehistoric rock art was investigated by Beck et al [107] with this type of XRF/XRD instrument. Gianoncelli et al [108] presented the design of this type of XRF/XRD instrument and the first test experiment of the instrument was implemented. The XRF/XRD instrument has the capability of characterizing the culture heritage objects with arbitrary shapes, especially suitable for being used in the case that the objects are non-moveable or have irregular surface or shapes. The complementary and accurate results on the two Egyptian murals dating from the New Kingdom were obtained by Pages-Camagna et al [109] with this type of XRF/XRD instrument. In 2014, Duran et al [110] used this type of XRF/XRD instrument to analyze an extremely valuable illuminated parchment. The XRF/XRD instrument played a crucial role during the analyses, and a complete and clear characterization was realized.

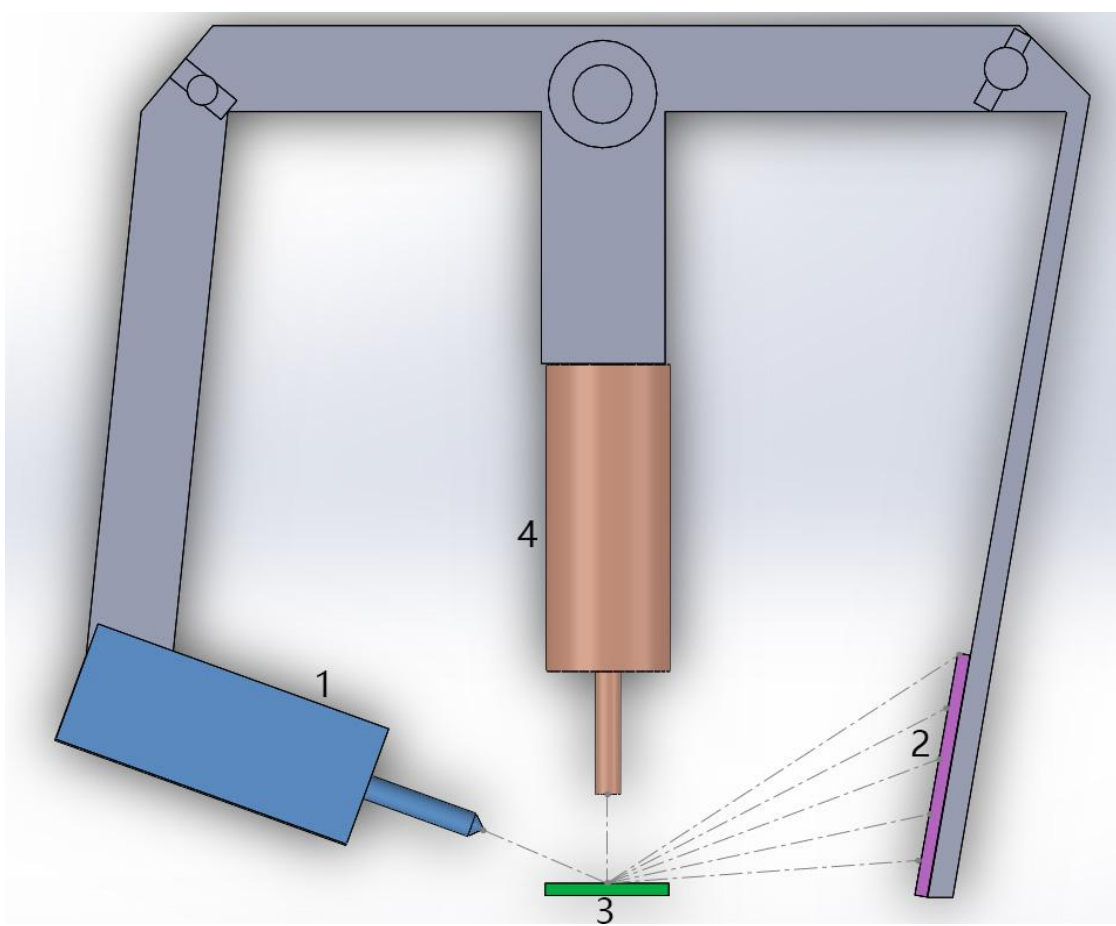

Fig. 5. The structural schematic diagram of fourth type of XRF/XRD instrument (1. X-ray tube, 2. IP detector, 3. Sample, 4. Si-PIN or SDD detector).

The fifth type of XRF/XRD instrument based on reflection configuration is composed of an X-ray tube and a CCD detector. As the image plate is not suitable for collecting the fluorescence and diffraction information simultaneously and need a data reader, only the CCD can be used as the two-dimensional detector for this type of XRF/XRD instrument. The structural schematic diagram is shown in Fig. 6. At present, such XRF/XRD instrument is becoming the most promising analytical tool in the research field of cultural heritage objects. There is no complicated structural design and no mechanical movements. Besides these advantages, such XRF/XRD instruments have more compact structure, faster data acquisition, easier data processing and interpretation, more accurate analytical results, smaller size and better portability. They almost have all the advantages of the XRF/XRD instruments aforementioned except for the smaller 2θ range. In fact, it is often enough to identify most of crystalline phases present in cultural heritage objects accurately. Therefore, the researchers are drawn more attention to such XRF/XRD instrument. Generally, the fluorescence and diffraction information can be collected by the CCD detector simultaneously. The two-dimensional diffraction image collected by the CCD detector can be converted to the conventional XRD pattern by the free software, i.e., FIT-2D. All components can be in fixed positions relative to each other to guarantee a stable and compact geometry. This type of instrument is very suitable for the investigation of cultural heritage objects in situ. For example, such an XRF/XRD instrument named as X-DUETTO was used by Sarrazin et al [111] for the study of cultural heritage objects. It was developed jointly by the Getty Conservation Institute and inXitu Inc. Most of technologies used in this XRF/XRD instrument came from the XRF/XRD instrument based on transmission configuration, i.e., CheMin [97] and Terra [99]. The X-DUETTO is composed of a miniature X-ray tube and a CCD detector and it can be used to investigate the large smooth objects such as murals in a nondestructive way. Some cultural heritage objects and mockup samples were investigated for testing the performance of the instrument [111]. The results corroborated the usefulness and accurateness of such XRF/XRD instrument. However, the authors [111] revealed that the limited clearance relative to the focusing plane was the primary limitation of the X-DUETTO instrument. Chiari et al [112] used the improved XRF/XRD instrument named as DUETTO to study the relationship between the displacement (clearance relative to the focusing plane) and peak shift (angular difference in 2θ from the standard value). The authors [112] derived the equation correlating the displacement versus peak shift, which was very useful for the data processing. In addition, the usefulness of the instrument was corroborated by the experiments in which the quantitative composition of alloy was determined, difference of the clad and electroplated daguerreo-types was distinguished and the beeswax used in painting was identified. Recently, the DUETTO2 has been developed and it was improved from the DUETTO [113]. The new DUETTO2 is smaller than the DUTTO, which makes it have better portability. The new software offers advanced controls and data processing with an intuitive, dynamic and user-friendly graphic interface. However, the specific application of the DUETTO2 for the study of cultural heritage objects has not been reported in literature presently.

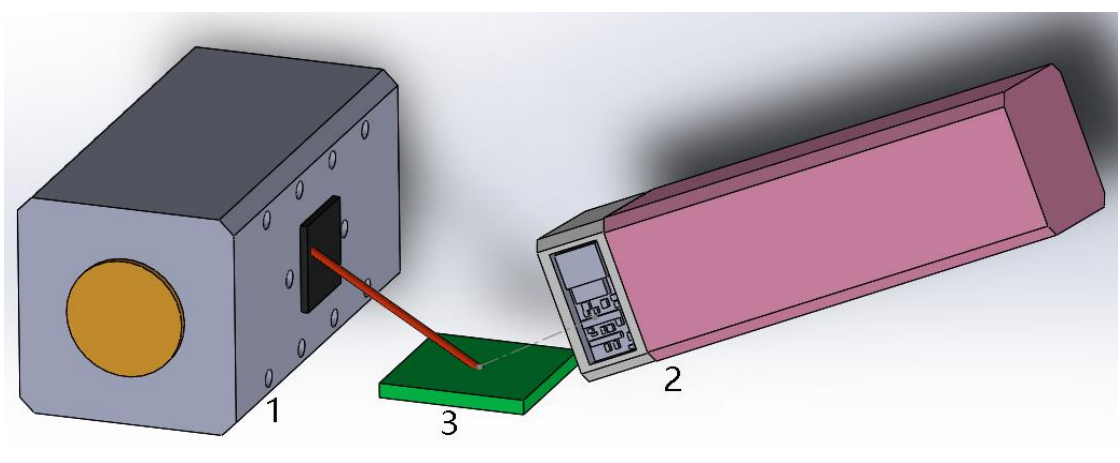

Fig. 6. The structural schematic diagram of fifth type of XRF/XRD instrument (1. X-ray tube, 2. CCD detector, 3. Sample).

**2.2 XRF/RS instruments**

Besides the XRF/XRD comprehensive instruments, the XRF/RS comprehensive instruments where the XRF and RS are integrated into a single device are also very useful for the investigation of cultural heritage objects. These XRF/RS comprehensive instruments mainly consist of the XRF and RS analytical systems. The XRF analytical system is composed of an X-ray tube and a Si-PIN or SDD detector, while the RS analytical system is mainly composed of a laser excitation source, a Raman probe and a spectrometer equipped with CCD. The laser lights are transmitted to the Raman probe and irradiate on the sample surface. Then, the Raman signals scattered by the sample are collected by the Raman probe and transmitted to the spectrometer. These XRF/RS comprehensive instruments are based on the reflection configuration, i.e., the X-ray tube, the detector and the Raman probe are on the same side of sample. The structural schematic diagram is shown in Fig. 7. The reflection configuration is beneficial to the nondestructive analyses of cultural heritage objects. The elemental information obtained by XRF not only provides a support to the structural information obtained by RS, but also gives additional information to the studied materials with low Raman signals. The complementary information provided by the two different types of analytical techniques contributes to the identification of complicated mixtures. For example, a remarkable paper on using a new mobile XRF/RS comprehensive instrument to investigate the Byzantine icons was presented by Andrikopoulos et al [84]. The instrument allowed comprehensive characterization of materials present in the culture heritage objects. Van Der Snickt et al [114] presented an impressive paper that has drawn attention deeply because that for the instrument, called PRAXIS, where the µ-RS and µ-XRF were integrated into a single device. The disadvantages of the µ-RS and the µ-XRF used separately were described by the authors [114] for explaining the necessity that two complementary techniques were integrated into a single device to demonstrate the importance of the XRF/RS comprehensive instrument for the nondestructive investigation of cultural heritage objects. The identification of pigments was carried out by Deneckere et al [115] using the fused Raman and XRF data, which showed the superiority and applicability of the combined use of RS and XRF from a view of data processing. In 2014, a portable XRF spectrometer coupled to a digital optical microscope and a Raman head for Raman measurement was developed by Guerra et al [89] for the in-situ analysis of culture heritage objects that were prohibited for sampling and moving. The potential of the XRF/RS instrument was confirmed by the characterization of various objects. As a matter of fact, the development and application of the XRF/RS comprehensive instrument also implied the development trend of analytical instruments in the research field of cultural heritage objects [89].

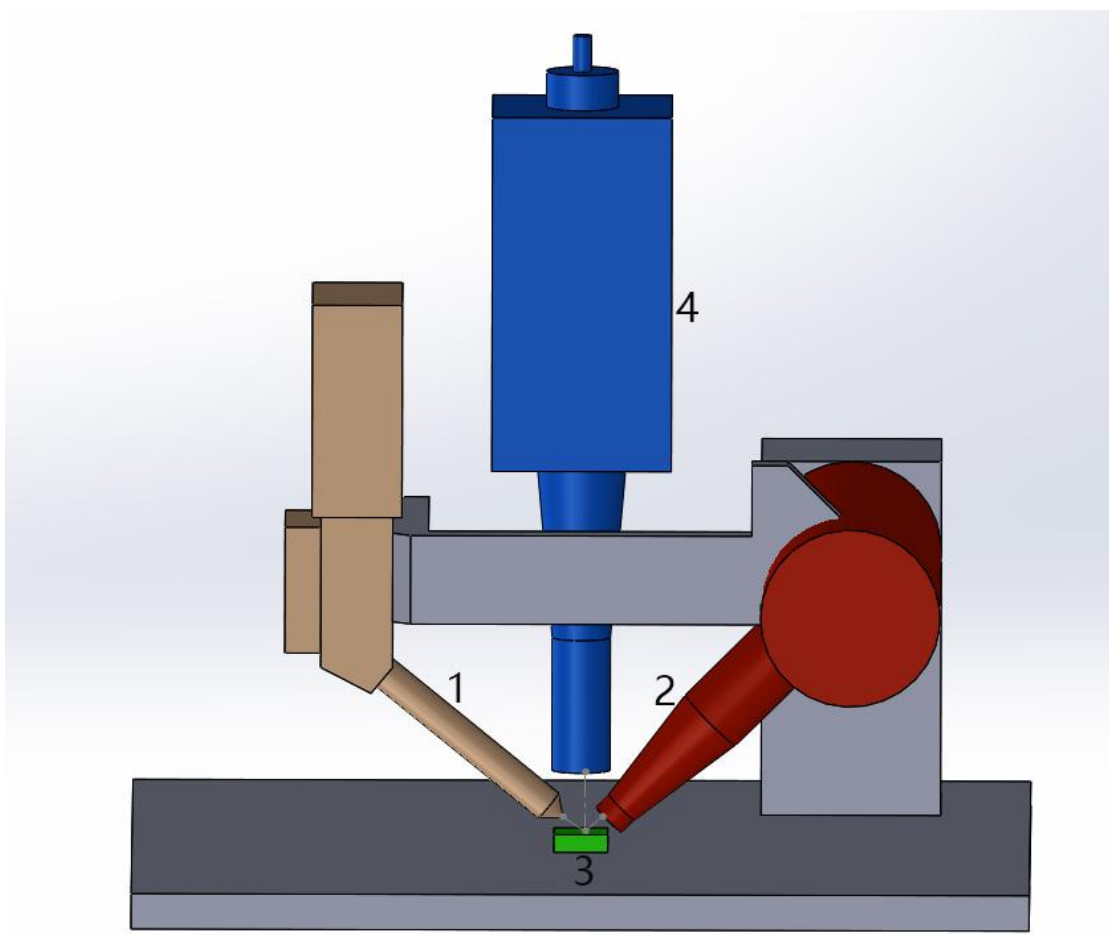

Fig. 7. The structural schematic diagram of XRF/RS comprehensive instrument (1. X-ray tube, 2. Si-PIN or SDD detector, 3. Sample, 4. Raman probe).

**2.3 XRF/XRD/RS comprehensive analytical instruments**

It is well known that different types of information on samples can be provided by various analytical techniques. Not all of analytical techniques are suitable for the study of cultural heritage objects due to their particularity, such as fragility, irreproducibility, significance and so on. The XRF, XRD and RS are often used to study cultural heritage objects. If more information can be obtained while analyzing these complicated cultural heritage objects, it would make the characterization of these objects more complete and accurate. The multi-technique method including different types of nondestructive analytical techniques is a choice for the investigation of some complicated cultural heritage objects that allow us to sample or transport. But for some precious objects that prohibit for sampling and transporting, a mobile comprehensive instrument may be more suitable for the nondestructive analyses. The XRF/XRD and XRF/RS comprehensive instruments have been widely used in the research field of the cultural heritage objects. For some cultural heritage objects, nor XRF/XRD or XRF/RS instrument can make the cultural heritage objects be characterized completely because only two different types of information cannot provide an unambiguous and complete identification. Hence, it is very necessary to use a comprehensive instrument that integrates more analytical techniques into a single device. The more analytical techniques are integrated into a single device, the richer the information is obtained. However, not all of analytical techniques can be integrated into a single device and it is very difficult to integrate more analytical techniques into a single instrument due to the limitations of components, light path, weight, size, mechanical structure and so on. This is why only two types of analytical techniques are currently integrated into a single device. With the development of electronics, detectors and so on, as well as the miniaturization of various components, the comprehensive analytical instruments that integrate more analytical techniques into a single device become possible. At present, there is not any comprehensive analytical instrument that integrates three different types of analytical techniques. Such comprehensive analytical instrument deserves to be developed because there is growing desire for more detailed information on cultural heritage objects. The selection of analytical techniques is crucial before developing such a comprehensive analytical instrument. Many aspects should be considered for the selection of different types of analytical techniques. Firstly, the feasibility should be considered. The analytical principles of different techniques are differently, therefore, ones should guarantee that the analytical techniques can be integrated into a single device. Meanwhile, these analytical techniques should be available easily. Secondly, the complementarity of different types of analytical techniques is very important for the investigation of cultural heritage objects that are composed of complicated mixtures. When the analytical results obtained by one type of analytical technique can provide the supplementary and auxiliary information for other analytical results, the unambiguous and complete identification of a complicated cultural heritage object will become easier. Thirdly, the nondestructive analytical techniques are preferred. After all, the cultural heritage objects are precious and irreproducible, and any destructive should be avoided. Fourthly, the portability of instrument should also be considered for a comprehensive instrument. The analytical techniques that cannot be provided by portable instrument cannot be selected to integrate into a comprehensive instrument. While the cultural heritage objects that prohibit for sampling and moving are studied in situ, the mobile comprehensive instrument can be transported to the field and provide the complete and accurate analytical results. Fifthly, the analytical techniques selected are able to be integrated into a single device from the point of view of structural design. Even though these analytical techniques have satisfied the five conditions aforementioned, it doesn't mean that they can be integrated into a single device successfully, because there are still some difficulties, such as the compatibility of different components, the complexity of mechanical structure design, the limitations of weight and size and so on. Therefore, to develop such a comprehensive instrument that integrates three different types of analytical techniques, it is essential that an innovative instrumental structure should be designed, and the smallest and advanced components should be used and so on. According to the selection principles and the analytical techniques that are widely used in the research field of cultural heritage objects, the XRF, XRD and RS are the best suitable analytical techniques that can be integrated into a single device. The elemental, crystalline and molecular structural information are complementary and they are crucial for better understanding cultural heritage objects. The XRF, XRD and RS analytical techniques are still essential for the accurate and complete characterization of the most of cultural heritage objects that are composed of complicated compositions. Therefore, a proposal that the XRF, XRD and RS analytical techniques are integrated into a single device where the elemental, crystalline and molecular structural information of the cultural heritage objects can be obtained simultaneously, is presented by us. The aim of the proposal is to develop a new, multifunctional, mobile and comprehensive XRF/XRD/RS analytical instrument for the complete, reliable and accurate identification of various culture heritage objects.

At present, we have successfully developed such a new XRF/XRD/RS comprehensive analytical instrument prototype. The structural schematic diagram of measurement head is shown in Fig. 8 and the three-dimensional (3D) diagram of XRF/XRD/RS comprehensive analytical instrument is shown in Fig. 9. The XRF/XRD/RS comprehensive analytical instrument prototype is shown in Fig. 10. It can be used to perform nondestructive analysis in situ. The prototype is composed of measurement head, 3D mobile platform and control system. The measurement head is composed of an X-ray tube (Moxtek, MAGNUM Reflection Source), a CCD detector (Andor, Newton SY) and a Raman probe module (Idea Optics, KUN). The control system contains the control software, control unit and a laptop. The control software was developed by our team. The prototype can perform scanning analysis. The Raman probe module is composed of a microscope objective, a Raman probe, a CCD camera and a connector. A 785-nm semiconductor laser was used and the maximum power is 500 mW. The target material of the X-ray tube is the copper and the power is 4 W. The active pixels of CCD detector are 1024×256. The key performance specifications of the XRF/XRD/RS comprehensive analytical instrument are as follows: XRF energy resolution is less than 230 eV, XRD 2θ is 21-49 ° and the angular resolution is approximately 0.5 °, the RS resolution is 6 cm$^{-1}$, the maximum scanning analysis area is 300×300 mm$^2$.

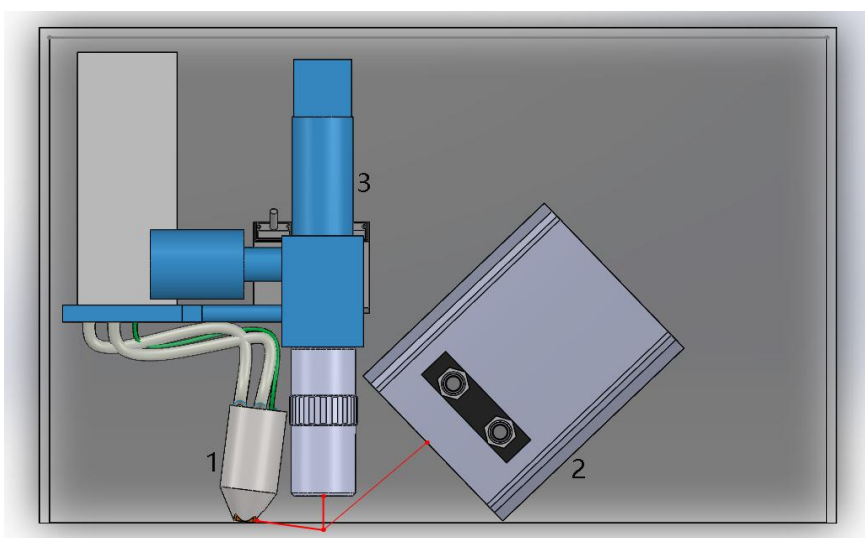

Fig. 8. The structural schematic diagram of measurement head (1. X-ray tube, 2. CCD detector, 3. Raman probe module).

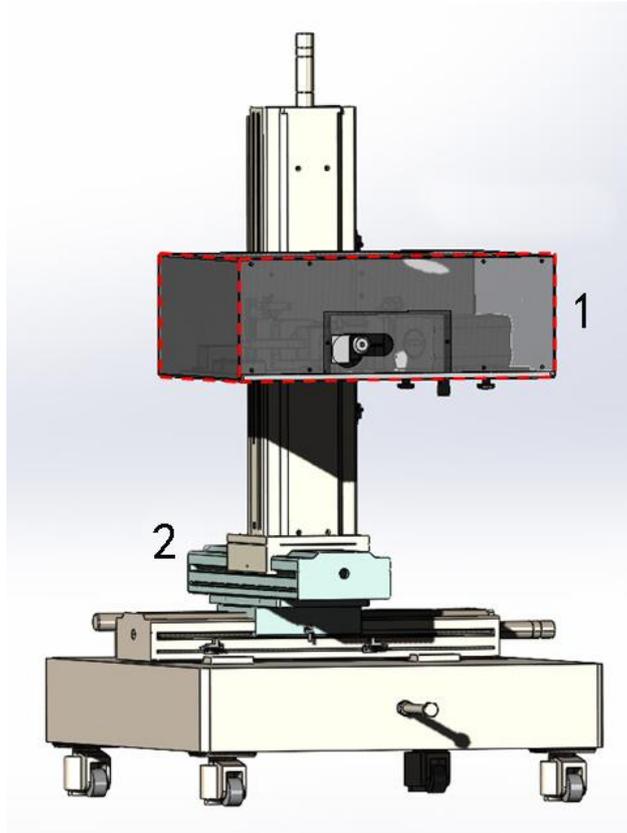

Fig. 9. The 3D diagram of XRF/XRD/RS comprehensive analytical instrument (1. Measurement head, 2. 3D mobile platform).

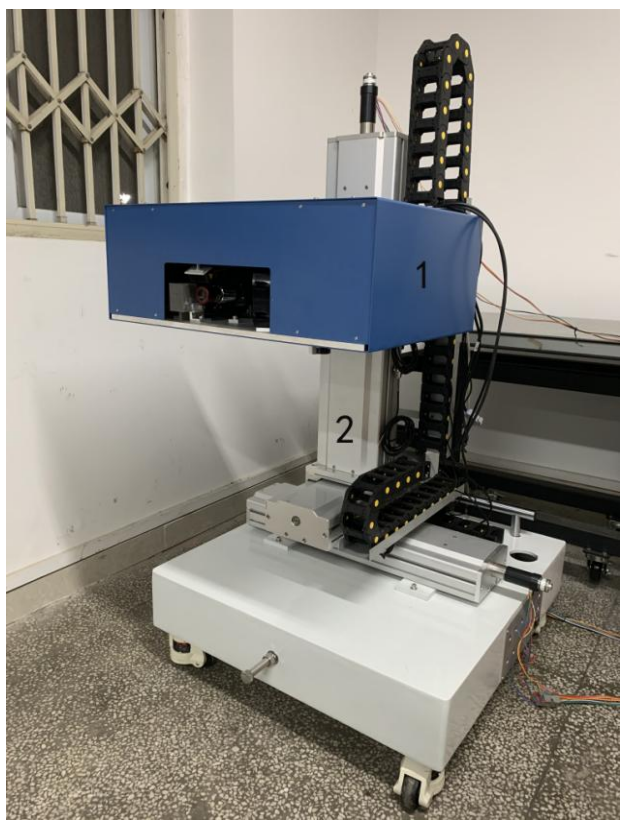

Fig. 10. The XRF/XRD/RS comprehensive analytical instrument prototype (1. Measurement head, 2. 3D mobile platform).

The XRF/XRD/RS comprehensive analytical instrument prototype has been used to analyze the zincite sample and several different color natural ore pigments for demonstrating the analysis performance and application potential, and the satisfactory results were obtained. As an example, the analysis results of zincite sample are shown in Figs. 11 to 14. The analysis results show that the zincite sample is mainly composed of the Zn and characterized by zincite (ZnO) phase, which is completely consistent with the known composition information.

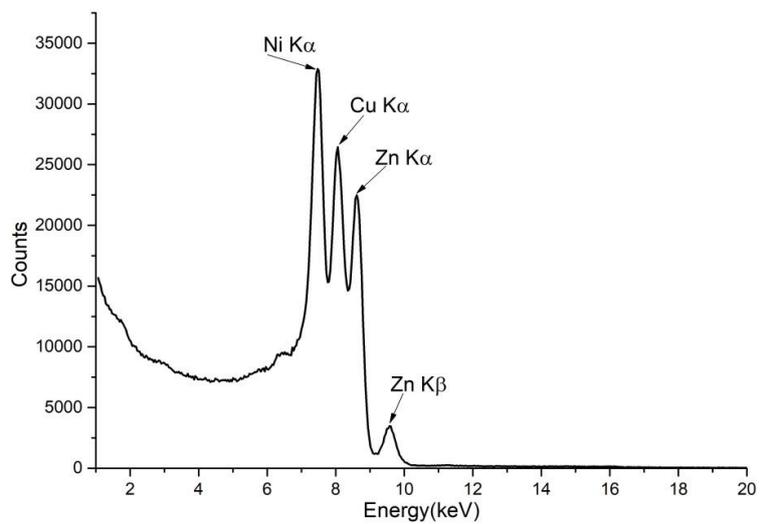

Fig. 11. The XRF spectrum of the zincite sample (the Ni peak comes from the nickel filter, while the Cu peak comes from the copper target material of the X-ray tube).

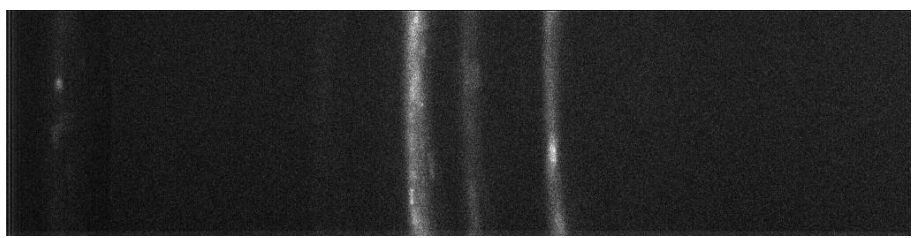

Fig. 12. The two-dimensional XRD pattern of the zincite sample on the CCD detector.

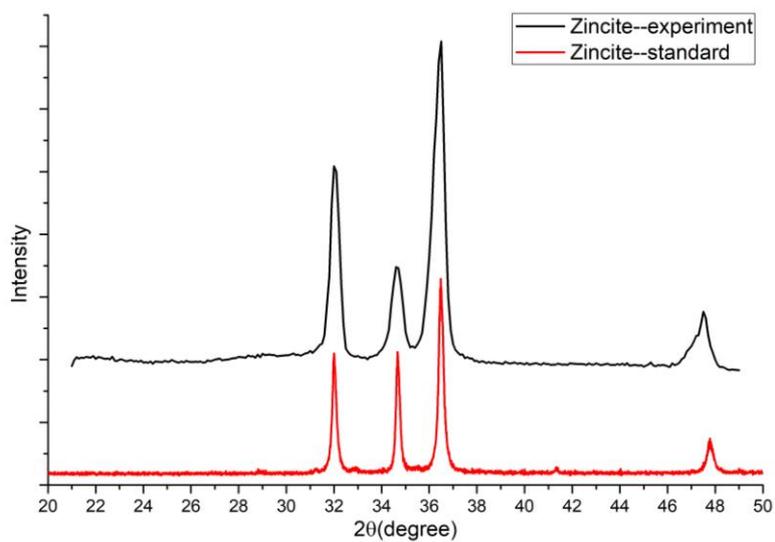

Fig. 13. The XRD spectrum of the zincite sample. The standard spectrum is taken from [116].

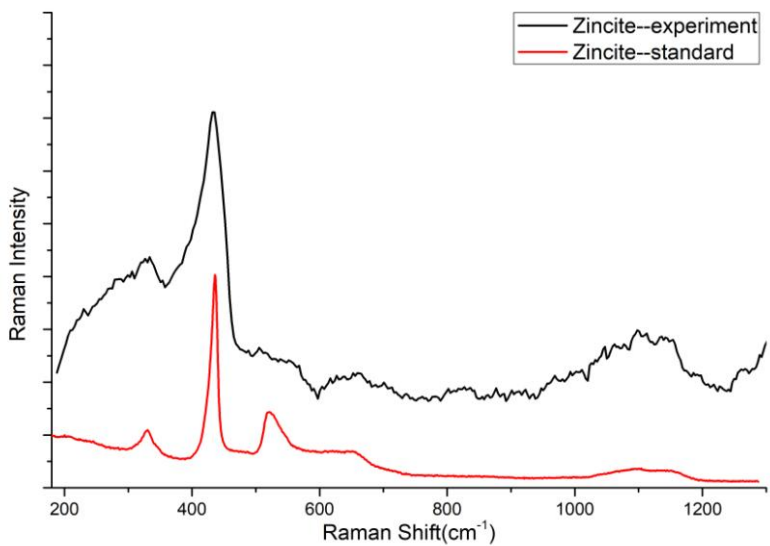

Fig. 14. The RS spectrum of the zincite sample. The standard spectrum is taken from [116].

Moreover, the XRF/XRD/RS comprehensive analytical instrument prototype has been used to analyze the yellow and white pigments board for demonstrating the scanning analysis performance, and the satisfactory results were obtained. The yellow (Orpiment, $As_2S_3$) and white (Dolomite, $CaMg(CO_3)_2$) pigments board is shown in Fig. 15(c). As an example, the elemental mapping results of yellow and white pigments board are shown in Fig. 15(a)-(b). The analysis results show that the As element is uniformly distributed in the blue dashed line box area in Fig. 15(a) corresponding to the blue dashed line box area in the Fig. 15(c), while the Ca element is uniformly distributed in the green dashed line box area in Fig. 15(b) corresponding to the green dashed line box area in the Fig. 15(c), which are consistent with the major elements in yellow and white pigments.

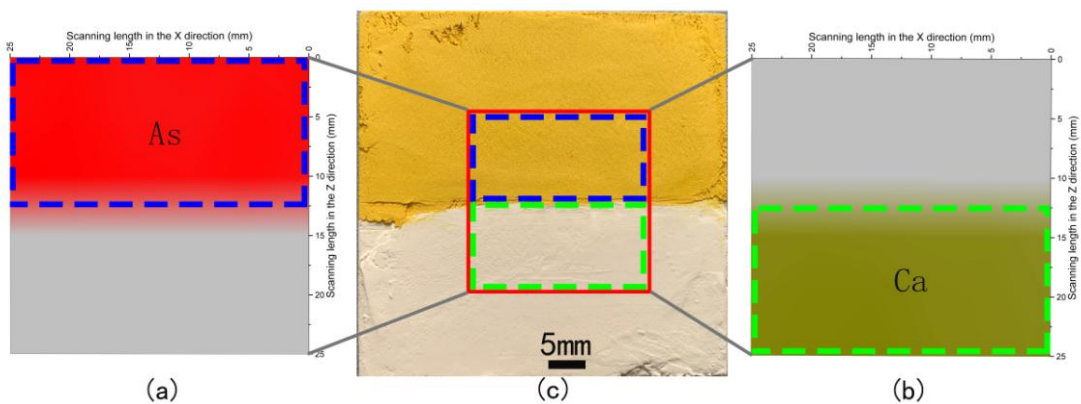

Fig. 15. The elemental mapping results of yellow and white pigments board (the gray areas in (a) and (b) indicate zero content).

Next, the XRF/XRD/RS comprehensive analytical instrument prototype will be improved and used to study the cultural heritage objects. We believe that the XRF/XRD/RS comprehensive analytical instrument will soon play an important role in the analyses of the cultural heritage objects that are composed of the complicated compositions. In addition, an interesting paper described a joint EU H2020 project known as SOLSA that is being implemented for developing an automatic expert system equipped with imaging, profilometer, sonic drilling, hyperspectral cameras and a combination of RS, XRF and XRD [117]. The main purpose of the SOLSA is to design and build a comprehensive analytical instrument that will be used to the geological and mining survey. The implementation of this project also implies the development trend for analysis instruments in different research fields for various materials characterization in the future. The SOLSA project and our new XRF/XRD/RS comprehensive analytical instrument prototype show that the comprehensive analytical instruments that integrate more analytical techniques into a single device are getting more and more attention. Some researchers have also shifted their focus to the study of such comprehensive analytical instruments [117].

## 3. Conclusions

With the growing interests for studying cultural heritage objects, ones desire to obtain more information on cultural heritage objects so that they can better understand and preserve these precious and irreproducible cultural heritage objects. A common instrument that integrated one type of analytical technique into a single device often cannot provide unambiguous and complete identification. Therefore, the multi-technique method with several individual instruments is used to study cultural heritage objects. Although the multi-technique method has many advantages and a large number of applications have demonstrated the usefulness, it also has some disadvantages. The comprehensive analytical instrument not only avoided these disadvantages, but also retained the advantages of the multi-technique method. The XRF/XRD and XRF/RS comprehensive analytical instruments that were widely used in the research field of cultural heritage objects were reviewed in detail, the XRF/XRD instruments were classified to different types according to the geometry structures of instruments, the type and the number of detectors. The advantages and disadvantages of each type of comprehensive analytical instruments are discussed. The comprehensive instruments where only two different types of analytical techniques are integrated into a single device cannot provide enough information for some complicated cultural heritage objects. It is necessary to develop a comprehensive instrument where several different types of analytical techniques are integrated into a single device. A new mobile XRF/XRD/RS comprehensive instrument prototype where the XRF, XRD and RS analytical techniques are integrated into a single device has been successfully developed by our team, and its preliminary application has proved its performance and application potential. It will play an important role in the research field of cultural heritage objects.

With the development of analytical technologies, advanced analytical instruments have become an important tool for art and archaeology researches. The more analytical techniques integrated into a single analytical instrument, the more sample information it can provide simultaneously at the same sample position. Complete and accurate sample information is beneficial to a better understanding of the samples under study, and powerful analytical instruments play a key role in promoting the further development of cultural heritage studies, archaeology, geology and other related research fields. It has been shown that the development of advanced comprehensive analytical instruments is obviously becoming the current trend.

**Data availability**

The datasets generated during and/or analyzed during the current study are available from the corresponding author on reasonable request.

**Abbreviations**

**XRF**: X-ray fluorescence

**XRD**: X-ray diffraction

**RS**: Raman spectrometry

**FTIR**: Fourier transform infrared spectroscopy

**EDXRF**: Energy dispersive X-ray fluorescence

**SEM-EDXS**: Scanning electron microscopy coupled to energy dispersive X-ray spectrometry

**OM**: Optical microscopy

**IRT**: Infrared thermography

**TEM**: Transmission electron microscope

**LIBS**: Laser-induced breakdown spectroscopy

**CCD**: Charge-coupled device

**FSI**: Front-side-illuminated

**BSI**: Back-side-illuminated

**SDD**: Silicon drift detector

**IP**: Imaging plate

**Acknowledgments**

The fund offered by Sichuan Science and Technology Program (project no. 2020ZDZX0004), China is appreciated.

**Funding**

This work was financially supported by Sichuan Science and Technology Program (project no. 2020ZDZX0004), China.

**Author contributions**

YJX studied different types of comprehensive analytical instruments for the investigation of cultural heritage objects and developed the XRF/XRD/RS comprehensive analytical instrument prototype where the XRF, XRD and RS analytical techniques are integrated into a single device and used the XRF/XRD/RS comprehensive analytical instrument prototype to analyze the samples and wrote the original paper draft. ZA supervised this study and studied different types of comprehensive analytical instruments for the investigation of cultural heritage objects and developed the XRF/XRD/RS comprehensive analytical instrument prototype and wrote the original draft paper and reviewed the paper. NH developed the XRF/XRD/RS comprehensive analytical instrument prototype and reviewed the paper and obtained the financial support. PW developed the XRF/XRD/RS comprehensive analytical instrument prototype and reviewed the paper. ZH developed the XRF/XRD/RS comprehensive analytical instrument prototype and offered some advices. ZHC drew some pictures and offered some advices. All authors read and approved the final manuscript.

**Competing interests**

The authors declare that they have no competing interests.

[4] I. Aliatis, D. Bersani, E. Campani, A. Casoli, P.P. Lottici, S. Mantovan, I-G. Marino, Pigments used in Roman wall paintings in the Vesuvian area. J. Raman Spectrosc. 41(11), 1537-1542 (2010).

[5] P. Holakooei, J.F. De Laperouse, M. Rugiadi, F. Caro, Early Islamic pigments at Nishapur, north-eastern Iran: studies on the painted fragments preserved at The Metropolitan Museum of Art. Archaeol Anthropol Sci. 10(1), 175–195 (2018).

[6] M. Vagnini, R. Vivani, E. Viscuso, M. Favazza, B.G. Brunetti, A. Sgamellotti, C. Miliani, Investigation on the process of lead white blackening by Raman spectroscopy, XRD and other methods: Study of Cimabue's paintings in Assisi. Vibrational Spectroscopy. 98, 41–49 (2018).

[7] M. Bakiler, B. Kırmızı, O.O. Ozturk, O.B. Hanyalı, E. Dag, E. Caglar, G. Koroglu, Material characterization of the Late Roman wall painting samples from Sinop Balatlar Church Complex in the black sea region of Turkey. Microchemical Journal. 126, 263–273 (2016).

[8] M.D. Robador, L. De viguerie, J.L. Perez-rodriguez, H. Rousseliere, P. Walter, J. Castaing, The structure and chemical composition of wall paintings from Islamic and Christian times in the Seville Alcazar. Archaeometry. 58(2), 255–270 (2016).

[9] L.D. Mateos, D. Cosano, D. Esquivel, S. Osuna, C. Jimenez-Sanchidrian, J.R. Ruiz, Use of Raman microspectroscopy to characterize wall paintings in Cerro de las Cabezas and the Roman villa of Priego de Cordoba (Spain). Vibrational Spectroscopy. 96, 143–149 (2018). [10] M.A. Legodi, D.D. Waal, Raman spectroscopic study of ancient South African domestic clay pottery. Spectrochim. Acta Part A. 66(1), 135–142 (2007).

[10] M.A. Legodi, D.D. Waal, Raman spectroscopic study of ancient South African domestic clay pottery. Spectrochim. Acta Part A. 66(1), 135–142 (2007).

[11] P. Colomban, B. Kırmızı, B. Zhao, J.B. Clais, Y. Yang, V. Droguet, Investigation of the Pigments
37

**Captions:**

Fig. 1. The structural schematic diagram of XRF/XRD instrument based on transmission configuration (1. X-ray tube, 2.Pinhole collimator, 3. Sample holder, 4. CCD detector).

Fig. 2. The structural schematic diagram of first type of XRF/XRD instrument (1. X-ray tube, 2. Si-PIN or SDD detector, 3. Sample, 4. Theta-Theta protractor).

Fig. 3. The structural schematic diagram of second type of XRF/XRD instrument (1. X-ray tube, 2. Si-PIN or SDD detector, 3. Sample).

Fig. 4. The structural schematic diagram of third type of XRF/XRD instrument (1. X-ray tube, 2. Two-dimensional CMOS detector, 3. Sample, 4. Si-PIN or SDD detector, 5. Theta-Theta

protractor).

Fig. 5. The structural schematic diagram of fourth type of XRF/XRD instrument (1. X-ray tube, 2. IP detector, 3. Sample, 4. Si-PIN or SDD detector).

Fig. 6. The structural schematic diagram of fifth type of XRF/XRD instrument (1. X-ray tube, 2. CCD detector, 3. Sample).

Fig. 7. The structural schematic diagram of XRF/RS comprehensive instrument (1. X-ray tube, 2. Si-PIN or SDD detector, 3. Sample, 4. Raman probe).

Fig. 8. The structural schematic diagram of measurement head (1. X-ray tube, 2. CCD detector, 3. Raman probe module).

Fig. 9. The 3D diagram of XRF/XRD/RS comprehensive analytical instrument (1. Measurement head, 2. 3D mobile platform).

Fig. 10. The XRF/XRD/RS comprehensive analytical instrument prototype (1. Measurement head, 2. 3D mobile platform).

Fig. 11. The XRF spectrum of the zincite sample (the Ni peak comes from the nickel filter, while the Cu peak comes from the copper target material of the X-ray tube).

Fig. 12. The two-dimensional XRD pattern of the zincite sample on the CCD detector.

Fig. 13. The XRD spectra of the zincite sample. The standard spectrum is taken from [116].

Fig. 14. The RS spectra of the zincite sample. The standard spectrum is taken from [116].

Fig. 15. The elemental mapping results of yellow and white pigments board (the gray areas in (a) and (b) indicate zero content).